\documentclass[aps,nofootinbib]{revtex4}
\usepackage{graphicx}
\begin{document}
\title{Black hole, string ball, and $p$-brane production at hadronic
supercolliders
\footnote{Talk presented in the SUSY02, DESY, Hamburg, Germany, June 2002}
}
\author{Kingman Cheung}
\affiliation{National Center for Theoretical Sciences, National Tsing Hua
University, Hsinchu, Taiwan, R.O.C.}

\begin{abstract}
In models of large extra dimensions, the string and Planck scales
become accessible at future colliders. When the energy scale is
above the string scale or Planck scale a number of interesting
phenomena occur, namely, production of stringy states, $p$-branes,
string balls, black hole, etc. In this Proceedings, we summarize
a recent work \cite{me0} on the production of black holes, string balls,
and $p$-branes at hadronic supercolliders, and discuss their
signatures. 
\end{abstract}
\maketitle

\section{Introduction}

In a model of large extra dimensions (ADD model) \cite{arkani},
the fundamental Planck scale can be as low as a few TeV, which
is made possible by localizing the SM particles on a
brane while gravity is free to propagate in all dimensions.
The observed Planck scale ($\sim 10^{19}$ GeV) is then a derived quantity.
Signatures for the ADD model can be divided into two categories:
sub-Planckian and trans-Planckian.  
The former has been studied extensively, while
the latter just recently received more attention, especially black hole
production in hadronic collisions.  

The fact that the fundamental Planck scale is
as low as TeV opens up an interesting possibility of
producing a large number of black holes at collider experiments
(e.g. LHC) \cite{hole,scott,greg}.  Reference \cite{emp} showed that a
BH localized on a brane will radiate mainly in the brane, instead
of radiating into the Kaluza-Klein states of gravitons of the
bulk. In this case, the BH so produced will decay mainly into the
SM particles, which can then be detected in the detector.  This
opportunity has enabled investigation of the properties of BH at
terrestrial collider experiments.

An important quantity of a BH is its entropy $S_{\rm BH}$.  To fulfill the
thermodynamical description, a BH requires a large entropy of
order of 25 \cite{giddings}.  Such an entropy requirement implies
that the BH mass must be at least five times the fundamental
Planck scale \cite{giddings,me}.  
This mass requirement makes the BH production not
as large as previously calculated in a number of works
\cite{scott,greg}, first pointed out in Ref. \cite{me}. In
addition, the signature of large multiplicity decay of a BH can only
happen when the entropy is large.  Even taking into account this
mass requirement, the event rate is still large enough for detection.  
A typical signature of the BH decay is a high
multiplicity, isothermal event, very much like a spherical
``fireball." 

Other interesting trans-Planckian phenomena include string
balls \cite{emparan} and  $p$-branes \cite{ahn}.
Dimopoulos and Emparan \cite{emparan} pointed out that when a BH reaches a
minimum mass, it transits into a state of highly excited and
jagged strings -- a string ball (SB). The transition point is at
\begin{equation}
M^{\rm min}_{\rm BH} = \frac{M_s}{g_s^2} \;,
\end{equation}
where $M_s$ is the string scale and $g_s$ is the string coupling.
Naively, SB's are stringy progenitors of BH's. The BH
correspondence principle states that properties of a BH with a
mass $M_{\rm BH}=M_s/g_s^2$ match those of a string ball 
with $M_{\rm SB}=M_s/g_s^2$. We can then equate the production
cross sections of SB and BH at the transition point. 
The existence of string balls could be
argued from the string point of view.  When the energy of the
scattering reaches the string scale, the scattering of particles is
no longer described by point-particle scattering but replaced by
string-string scattering.  As the energy goes further up, the
strings become highly excited, jagged and entangled string states,
and become like a string ball.   When the energy reaches the
transition point, it turns into a BH.  In the above
we mentioned a large entropy requirement on the BH
in order for the object to be a BH.  Such a large mass requirement
makes the production cross section smaller than previously
thought.  Here in the case of SB's, the mass requirement is
substantially lower, thus the production rate is significantly
higher.  Hence, an SB is more interesting in the experimental point
of view if it decays with a distinct signature. Dimopoulos and
Emparan \cite{emparan} argued that the decay of a SB is similar to
that of a BH, i.e., a high multiplicity decay into the SM
particles, though in some intermediate stages the SB decays more
likely into bulk particles.

A $p$-brane is a solution to the Einstein equation in higher dimensions
\cite{ahn}. A BH can be considered a $0$-brane.  In
fact, the properties of $p$-branes reduce to those of a BH in the
limit $p\to 0$. In extra dimension models, in which there are
large extra dimensions and small extra dimensions of the size of the 
Planck length, let a $p$-brane wrap on $r$ small and $p-r$ large
dimensions. It was found \cite{ahn} that the production of
$p$-branes is comparable to BH's only when $r=p$, i.e., the
$p$-brane wraps entirely  on the small dimensions only.  If $r< p$, 
the production of $p$-branes would be suppressed by
powers of $(M_*/M_{\rm Pl})$, where $M_*$ is the fundamental scale of the
$4+n$ dimensions.  Therefore, here we only consider
the case in which $r=p$. The decay of $p$ branes is not well
understood. One interesting possibility is cascade into branes of
lower dimensions until they reach the dimension of zero.  Whether
the zero brane is stable depends on the model.  Another
possibility is the decay into brane and bulk particles, thus
experimentally the decay can be observed.  Or it can be a combination
of cascade into lower-dimensional branes and direct decays.
$p$-brane production in the unconfined scenario (e.g. in universal extra
dimensions or fat-brane scenario) was studied in Ref. \cite{chou}.

BH production has also been studied in cosmic ray experiments and neutrino
telescopes \cite{feng}, which are complementary to or even better than 
the LHC for BH discovery.

\section{Production}

\subsection{Black holes}

A black hole is characterized by its Schwarzschild radius $R_{\rm BH}$ and
its entropy, which
depend on the mass $M_{\rm BH}$ of the BH.  A simplified picture for
BH production is as
follows.  When the colliding partons have a center-of-mass (c.o.m.) energy
above some thresholds of order of the
Planck mass and the {\it impact parameter} less
than the Schwarzschild radius $R_{\rm BH}$, a BH is formed
and almost at rest in the c.o.m. frame.  The BH so produced will decay
thermally (regardless of the incoming particles) and thus isotropically
in that frame.

The Schwarzschild radius $R_{\rm BH}$ of a BH of mass $M_{\rm BH}$ in
$4+n$ dimensions is given by \cite{myers}
\begin{equation}
\label{r}
R_{\rm BH} = \frac{1}{M_D}\; \left (
\frac{M_{\rm BH}}{M_D} \right)^{\frac{1}{n+1}}\;
\left( \frac{ 2^n \pi^{ \frac{n-3}{2}} \Gamma(\frac{n+3}{2} )}{n+2}
\right )^{\frac{1}{n+1}}
= \frac{1}{M_D}\; \left (
\frac{M_{\rm BH}}{M_D} \right)^{\frac{1}{n+1}}\; f(n)
\;,
\end{equation}
where $f(n)$ is introduced for convenience and
 $M_D$ is the fundamental Planck scale.
The radius $R_{\rm BH}$ is much smaller than the size of the extra dimensions.
When the colliding partons with a center-of-mass
energy $\sqrt{\hat s} \agt M_{\rm BH}$ pass within a distance less than
$R_{\rm BH}$,  a black hole of mass $M_{\rm BH}$ is formed and the rest of
the energy, if there is any, is radiated as ordinary SM particles \cite{me}.
This semiclassical argument calls for a geometric approximation 
for the cross section for producing a BH of mass $M_{\rm BH}$ as
\begin{equation}
\label{geo}
\sigma(M_{\rm BH}^2 ) \approx \pi R_{\rm BH}^2 \;.
\end{equation}
In the $2\to 1$ subprocess, the c.o.m. energy of the colliding partons is just
the same as the mass of the BH, i.e., $\sqrt{\hat s}=M_{\rm BH}$, which
implies a subprocess cross section
\begin{equation}
\label{2to1}
\hat \sigma(\hat s) = \int \; d\left(\frac{M_{\rm BH}^2}{\hat s}\right )\;
\pi R_{\rm BH}^2 \; \delta\left( 1 - M^2_{\rm BH}/\hat s \right )
= \pi R_{\rm BH}^2 \;.
\end{equation}
On the other hand, for the $2\to k (k\ge 2)$ subprocesses the subprocess
cross section is
\begin{equation}
\label{2to2}
\hat \sigma(\hat s) = \int^{1}_{ (M^2_{\rm BH})_{\rm min}/\hat s} \;
d\left(\frac{M_{\rm BH}^2}{\hat s}\right )\; \pi R_{\rm BH}^2 \;.
\end{equation}

The entropy of a BH is given by
\cite{myers}
\begin{equation}
\label{entropy}
S_{\rm BH} = \frac{4\pi}{n+2}\; \left (
\frac{M_{\rm BH}}{M_D} \right)^{\frac{n+2}{n+1}}\;
\left( \frac{ 2^n \pi^{ \frac{n-3}{2}} \Gamma(\frac{n+3}{2} )}{n+2}
\right )^{\frac{1}{n+1}} \;.
\end{equation}
 To ensure the validity of the above
classical description of a BH \cite{giddings}, the entropy must be
sufficiently large, of order 25 or so.   We verified 
that when $M_{\rm BH}/M_D \agt 5$, the entropy $S_{\rm BH}\agt
25$ \cite{me0}.  Therefore, to avoid getting into the nonperturbative regime
of the BH and to ensure the validity of the semiclassical
formula, we restrict the mass of the BH to be $M_{\rm BH} \ge 5 M_D$.

\subsection{String balls}

According to the BH correspondence principle, the production cross
section of a string ball or a BH should be smoothly joined at $M_{\rm BH}=
M_s/g_s^2$, i.e.,
\[
\left. \sigma(SB) \right|_{M_{SB} = M_s/g_s^2} =
\left. \sigma(BH) \right|_{M_{BH} = M_s/g_s^2} \;.
\]
The production cross section for string balls with mass 
between the string scale $M_s$
and $M_s/g_s$ grows with $s$ until $M_s/g_s$, beyond which, due to unitarity,
it should stay constant.  Therefore, we can use the BH cross section and match
to the string ball cross section at the transition point $M_s/g_s^2$.
This string ball cross section then stays constant between $M_s/g_s$ and
$M_s/g_s^2$.  Then below $M_s/g_s$ the string ball cross section grows like
$M_{\rm SB}^2/M_s^4$.

The cross sections for the SB or BH are given by
\begin{equation}
\hat\sigma ({\rm SB/BH}) = \left \{ \begin{array}{ll}
\frac{\pi}{M_D^2}\; \left (
\frac{M_{\rm BH}}{M_D} \right)^{\frac{2}{n+1}}  \left[ f(n) \right ]^2 &
                         \frac{M_s}{g_s^2} \le M_{\rm BH} \\
\frac{\pi}{M_D^2}\; \left (
\frac{M_s/g_s^2}{M_D} \right)^{\frac{2}{n+1}}  \left[ f(n) \right ]^2
 = \frac{ \pi}{M_s^2}  \left[ f(n) \right ]^2 &
                \frac{M_s}{g_s} \le M_{\rm SB} \le \frac{M_s}{g_s^2} \\
\frac{ \pi g_s^2 M_{\rm SB}^2 }{M_s^4}  \left[ f(n) \right ]^2 &
                M_s \ll M_{\rm SB} \le \frac{M_s}{g_s}
\end{array}
 \right. \;,
\end{equation}
in which we have set $M_D^{n+2} = \frac{M_s^{n+2}}{g_s^2 }$.

\subsection{$p$-Branes}

Consider an uncharged and static $p$-brane with a mass $M_{p \rm B}$ in 
$(4+n)$-dimensional 
space-time ($m$ small Planckian size and $n-m$ large size extra
dimensions such that $n \ge p$).
Suppose the $p$-brane wraps on $r (\le m)$ small extra dimensions and on
$p-r (\le n-m)$ large extra dimensions.
Then the ``radius'' of the $p$-brane is
\begin{equation}
\label{rp}
R_{p \rm B} = \frac{1}{\sqrt{\pi} M_*} \, \gamma(n,p) \,
V_{p\rm B}^{ \frac{-1}{1+n-p} } \,
\left( \frac{M_{p\rm B}}{M_*} \right)^{ \frac{1}{1+n-p} } \;,
\end{equation}
where $V_{p\rm B}$ is the volume wrapped by the $p$-brane in units of the
Planckian length.
The $M_*$ is related to $M_D$ by
\[
M_D^{n+2} = \frac{(2\pi)^n}{8 \pi}\, M^{n+2}_* \;.
\]  
Recall $M_{\rm Pl}^2 = M_*^2 l_{n-m}^{n-m} l_m^m$,
where $l_{n-m}\equiv L_{n-m}\,M_{*}$ and $l_m \equiv L_m\,M_{*}$ 
are the lengths of the size of the large and small
extra dimensions in units of Planckian length ($\sim 1/M_*$).  Then
$V_{p\rm B}$ is given by
\begin{equation}\label{vp}
V_{p \rm B} = l_{n-m}^{p-r} \, l_m^{r} \approx \left( \frac{M_{\rm Pl}}{M_*}
\right )^{ \frac{2(p-r)}{n-m} } \;,
\end{equation}
where we have taken $l_m \equiv L_m \,M_{*} \sim 1$.  
The function $\gamma(n,p)$ is given by
\begin{equation}\label{gamma}
\gamma(n,p) = \left[ 8 \Gamma \left( \frac{3+n-p}{2} \right) \sqrt{
\frac{1+p}{(n+2)(2+n-p)} } \right ]^{\frac{1}{1+n-p} } \;.
\end{equation}
The $R_{p\rm B}$ reduces to the $R_{\rm BH}$ in the limit $p=0$.

The production cross section of a $p$-brane is similar to that of BH's, based
on a naive geometric argument \cite{ahn}, i.e.,
\begin{equation}
\hat \sigma (M_{p\rm B}) = \pi R^2_{p\rm B} \;.
\end{equation}
Therefore, the production cross section for a $p$-brane is the same as BH's
in the limit $p=0$ (i.e., a BH can be considered a $0$-brane).
In $2\to1$ and $2\to k \;(k\ge 2)$ processes, the parton-level cross sections
are given by similar expressions in Eqs. (\ref{2to1}) and (\ref{2to2}),
respectively.
Assuming that their masses are the same and the production threshold
$M^{\rm min}$ is the same, the ratio of cross sections is
\begin{equation}
\label{R}
R \equiv
\frac{\hat \sigma (M_{p\rm B}=M)}{\hat \sigma (M_{\rm BH}=M)}
= \left ( \frac{M_*}{M_{\rm Pl}} \right)^{\frac{4(p-r)}{(n-m)(1+n-p)}}
\, \left(\frac{M}{M_*} \right )^{ \frac{2p}{(1+n)(1+n-p)} } \,
\left( \frac{\gamma(n,p)}{\gamma(n,0)} \right )^2 \;.
\end{equation}
In the above equation, the most severe suppression factor is in the first
set of parentheses on the right-hand side.  Since we are considering physics of
TeV $M_*$, the factor $(M_*/M_{\rm Pl}) \sim 10^{-16}-10^{-15}$.  
Thus, the only
meaningful production of a 
$p$-brane occurs for $r=p$, and then their production
is comparable.  

\section{Production at the LHC and VLHC}

The production of BH's and SB's depends on $M_s,n,M_D,g_s$. 
Since we also require the
transition point ($M_s/g_s^2$) at $5M_D$, we can therefore solve
for $M_s$ and $g_s$ for a given pair of $M_D$ and $n$.  
We present the
results in terms of $M_D$ and $n$.  The minimum mass requirement
for the SB is set at $2M_s$.  The production of a $p$-brane also
depends on $m$ and $r$.  For an interesting level of event rates,
$r$ has to be equal to $p$, i.e., the $p$-brane wraps entirely on
small (of Planck length) extra dimensions.  

\begin{figure}[t!]
\includegraphics[width=7.95cm]{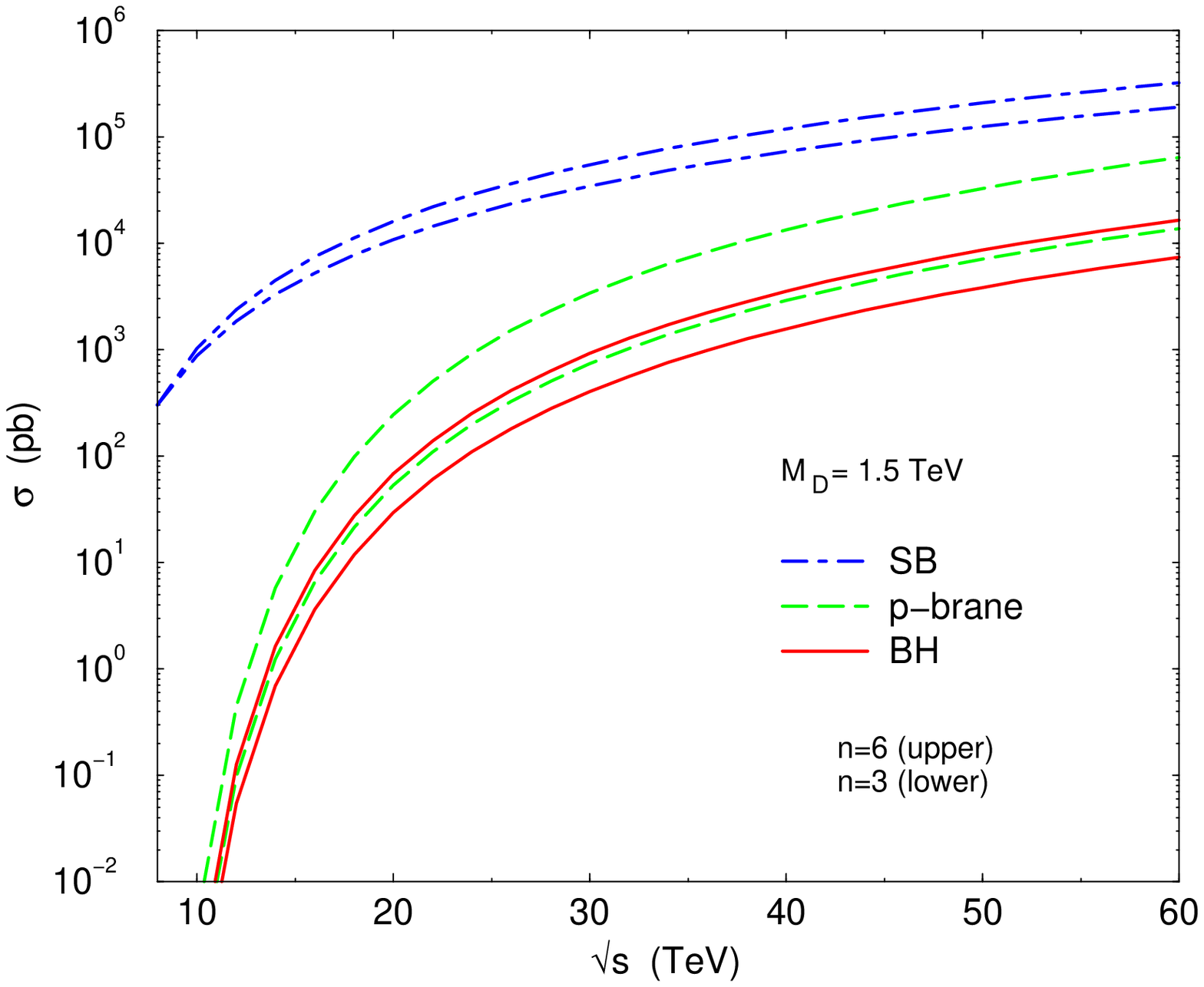}
\includegraphics[width=7.95cm]{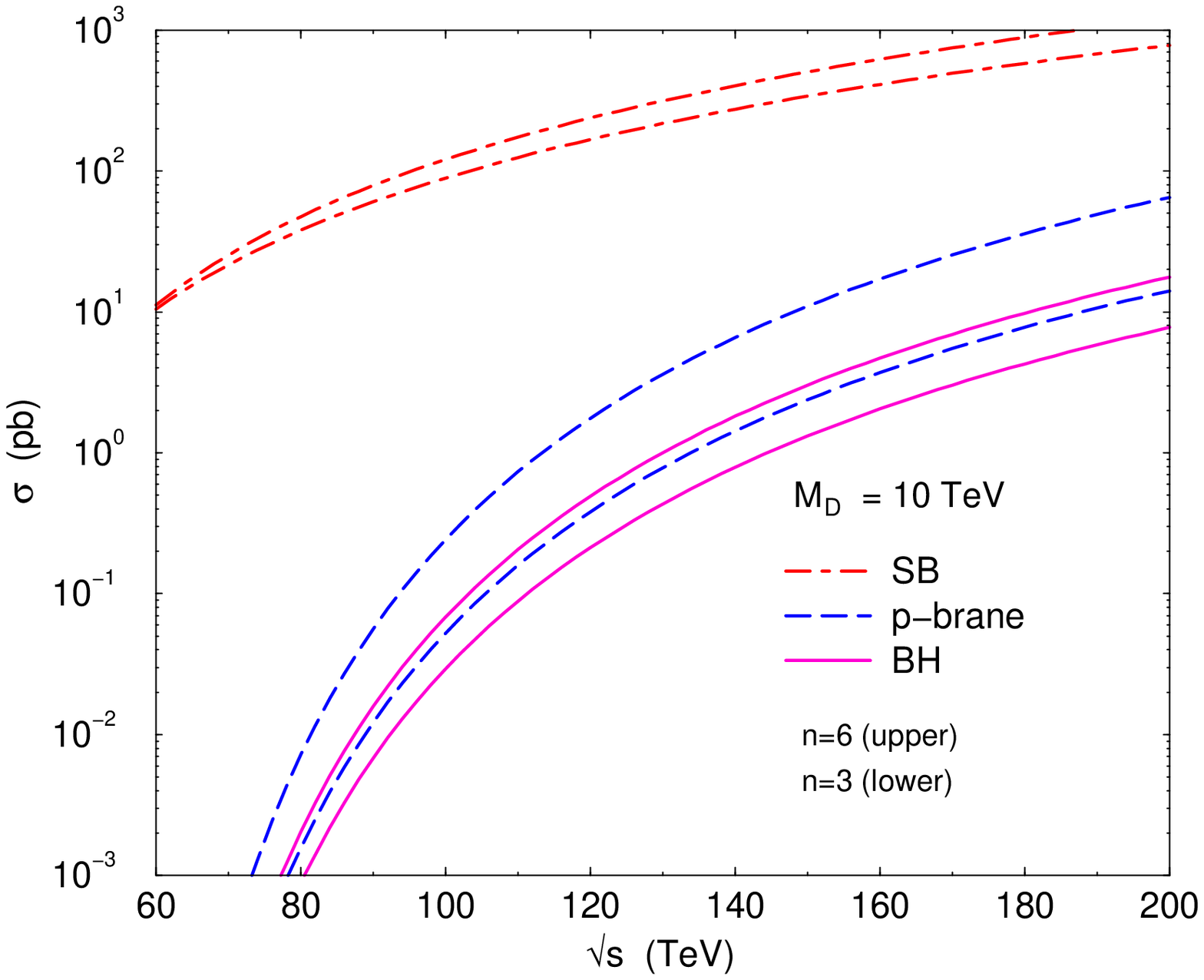}
\caption{\small
\label{total}
Total production cross section $\sigma$ for black hole (BH), 
string ball (SB), and $p$-brane ($p$B) at $pp$
collisions vs $\sqrt{s}$ for $n=3$ and $6$.
A fundamental scale (a) $M_D=1.5$ TeV and (b) $M_D=10$ TeV is used.
The minimum mass
on the BH and $p$-brane is $M_{\rm BH}^{\rm min}, M_{p\rm B}^{\rm min}=5 M_D$,
while that on SB is $M_{\rm SB}^{\rm min}=2 M_s$.  
}
\end{figure}

In Fig. \ref{total}(a), we show the total production cross sections for
BH's, SB's, and $p$-branes, including the $2\to 1$ and $2\to 2$
subprocesses (when computing the $2\to 2$ subprocess we require a
$p_T$ cut of 500 GeV to prevent double counting).  Typically, the
$2\to 2$ subprocess contributes at a level of less than 10\%.  For the
BH, SB, and $p$-brane, we show the results for $n=3$ and $n=6$. The
results for $n=4,5$ lie in between.  Since we require $M_{\rm
BH}^{\rm min}, \, M^{\rm min}_{p\rm B}=5 M_D$, their production is
only sizable when $\sqrt{s}$ reaches about 10 TeV, unlike the SB,
which only requires $M_{\rm SB}^{\rm min}=2M_s$. The $p$-brane
cross section is about a few times larger than the BH, as we have
chosen $r=p=m=n-2$.  String ball production is, on average, two
orders of magnitude larger than that of a BH in the energy range
between 20 and 60 TeV.  Below 20 TeV (e.g., at the LHC), the SB
cross section is at least three orders of magnitude larger than the
BH.
Integrated cross sections for the LHC in Table
\ref{table-lhc}.
Sensitivity information can be drawn from the table.  The event rates
for BH and $p$-brane production are negligible for $M_D=2.5$ TeV and only
moderate at $M_D=2$ TeV.  At $M_D=2$ TeV, the number of BH events that
can be produced in one year running (100 fb$^{-1}$) is about $120-340$
for $n=3-7$ while the number for $p$-brane events is $210-1300$.  Therefore,
the sensitivity for a detectable signal rate for a BH and a $p$-brane is
only around 2 TeV, if not much larger than 2 TeV.
The SB event rate is much higher.  Even at $M_D=3$ TeV, the cross section
is of order of 30 pb.  In Table \ref{table-lhc}, we also show the 
$\sigma({\rm SB})$ for $M_D=4-6$ TeV.  Roughly, the sensitivity is around 
6 TeV.

The VLHC (very large hadron collider) is another $pp$ accelerator
under discussions \cite{vlhc} in the Snowmass 2001 \cite{snowmass}.
The preliminary plan is to have an initial stage of about 40--60
TeV center-of-mass energy, and later an increase up to 200 TeV.  
The targeted luminosity is $(1-2)\times 10^{34}\,{\rm cm}^{-2} {\rm s}^{-1}$. 
In Fig. \ref{total}(b), we show the total production cross sections for
BH's, SB's, and $p$-branes for $\sqrt{s}=60-200$ TeV and for $n=3$ and $6$.
We found that the sensitivity reaches for BH and $p$-brane production
are roughly between $6$ and $7$ TeV for $\sqrt{s}=50$ TeV,
$10$ and $13$ TeV for $\sqrt{s}=100$ TeV, 
$14$ and $18$ TeV for $\sqrt{s}=150$ TeV,
and $20$ and $25$ TeV for $\sqrt{s}=200$ TeV.  These estimates are rather crude
based on the requirement that the number of raw events is $\agt 50-100$.

\begin{table}[th!]
\caption{\small \label{table-lhc}
Total cross sections in pb for the production of BH, SB, and $p$-brane,
for various values of $n$ and $M_D$ at the LHC.
The minimum mass
on the BH and $p$-brane is $M_{\rm BH}^{\rm min}, M_{p\rm B}^{\rm min}=5 M_D$,
while that on SB is $M_{\rm SB}^{\rm min}=2 M_s$.
}
\medskip
\begin{ruledtabular}
\begin{tabular}{clll}
                &  $n=3$ &  $n=5$ &  $n=7$ \\
\hline
            \multicolumn{4}{c}{\underline{BH}} \\
\underline{$M_D$ (TeV)}     &        &        &   \\
  1.5     &  $0.70$              & $1.3$ & $1.9$            \\
  2.0     &  $1.2\times 10^{-3}$ & $2.2\times10^{-3}$ & $3.4\times10^{-3}$\\
  2.5     &  $1.3\times 10^{-8}$ & $2.4\times10^{-8}$ & $3.6\times10^{-8}$\\
  \multicolumn{4}{c}{\underline{SB}} \\
\underline{$M_D$ (TeV)}     &        &        &         \\
  1.5     &  $3300$  & $4100$ & $4900$            \\
  2.0     &  $590$   & $670 $ & $760 $  \\
  2.5     &  $130$   & $130 $ & $140$  \\
  3.0     &  $33$    & $29  $ & $28$  \\
  4.0     &  $2.4$    & $1.5  $ & $1.1$  \\
  5.0     &  $0.16$    & $0.060  $ & $0.033$  \\
  6.0     &  $0.0091$    & $0.0015  $ & $0.00044$  \\
  \multicolumn{4}{c}{\underline{$p$-brane}} \\
\underline{$M_D$  (TeV)}     &        &        &         \\
  1.5     &  $1.2$  & $4.0$ & $7.6$            \\
  2.0     &  $2.1\times10^{-3}$   & $6.9\times10^{-3}$ & $0.013$  \\
  2.5     &  $2.3\times10^{-8}$   & $7.3\times10^{-8}$ & $1.4\times10^{-7}$
\end{tabular}
\end{ruledtabular}
\end{table}

\section{Decay Signatures}

The main phase of the decay of a BH is via the Hawking evaporation. 
An important observation is that the wavelength $\lambda$
of the thermal spectrum corresponding to 
the Hawking temperature is larger than the size of the BH.
This implies that the BH evaporates like a point source in $s$-waves,
therefore it decays equally into brane and bulk modes, and will not
see the higher angular momentum states available in the extra
dimensions.  Since on the brane there are many more particles than
in the bulk,  the BH decays dominantly into brane modes,
i.e., the SM particles in the setup.
Furthermore, the BH evaporates ``blindly" into all degrees of freedom.
The ratio of the degrees of freedom for gauge bosons, quarks, and leptons is
$29:72:18$ (the Higgs boson is not included).  Since the $W$ and $Z$ decay
with a branching ratio of about 70\% into quarks,
 and the gluon also gives rise to hadronic activities, the final ratio
of hadronic to leptonic activities in the BH decay is about $5:1$
\cite{scott}.

Another important property of the BH decay is the large number
of particles, in accord with the large entropy in Eq.
(\ref{entropy}), in the process of evaporation. It was shown
\cite{scott,greg} that the average multiplicity $\langle N
\rangle$ in the decay of a BH is order of $10-30$ for $M_{\rm BH}$
being a few times $M_D$ for $n=2-6$. Since we are considering
the BH that has an entropy of order 25 or more, it guarantees a
high multiplicity BH decay. The BH decays more or less
isotropically and each decay particle has an average energy of a
few hundred GeV.  Therefore, if the BH is at rest, the event is
very much like a spherical event with many particles of hundreds
of GeV pointing back to the interaction point (very much like a 
fireball).
On the other hand, if the BH is produced in
association with other SM particles (as in a $2\to k$ subprocess),
the BH decay will be a boosted spherical event on one side (a boosted
fireball), the
transverse momentum of  which is balanced by a few
particles on the other side \cite{me}.
Such spectacular events should have a negligible background.

Highly excited long strings emit massless quanta
with a thermal spectrum at the {\it Hagedorn} temperature.
At $M_{\rm SB} \alt M_s/g_s^2$, the wavelength $\lambda$ corresponding to
the thermal spectrum at the Hagedorn temperature
is larger than $R_{\rm SB}$.
This argument is very similar to that of the BH, and so
the string ball radiates like a point source and emits in $s$-waves equally
into brane and bulk modes.  With many more particles (SM particles)
on the brane than in the bulk, the SB radiates mainly into the SM
particles.
When $M_{\rm SB}$ goes below $M_s/g_s^2$, the SB has the tendency
to puff up to a {\it random-walk size} as large as the $\lambda$ of
the emissions \cite{emparan}.
  Therefore, it will see more of the higher angular momentum states
available in the extra dimensions.
Thus, it decays more into the bulk modes, but it is only temporary.
When the SB decays further, it shrinks back to the string size
 and emits as a point source again \cite{emparan}.
Most of the time the SB decays into SM particles.
On average, a SB decays into invisible quanta somewhat more often than
a BH does.

The decay of $p$-branes is not well understood, to some extent we do not
even know whether it decays or is stable.
Nevertheless, if it decays one
possibility is the decay into lower-dimensional branes, thus leading to
a cascade of branes.  Therefore, they eventually decay to a number of
 $0$-branes, i.e., BH-like objects.
This is complicated by the fact that
 when the $p$-branes decay, their masses 
might not be high enough to become BH's.
 Therefore, the final $0$-branes might be some excited string states or string
 balls. Whether the zero brane is stable or not depends on models.  Another
possibility is decay into brane and bulk particles, thus
experimentally the decay can be observed.  Or it can be a combination
of cascade into lower-dimensional branes and direct decays.
Since the size $R_{p\rm B}$ is much smaller than the size of the large
extra dimensions, we expect $p$-branes to decay mainly into brane particles.
However, the above is quite speculative.

This research was supported in part by the National Center
for Theoretical Science under a grant from the National Science
Council of Taiwan R.O.C.


\end{document}